\documentclass[%
 reprint,
amsmath,
amssymb,
pra,
floatfix,
balancelastpage,
]{revtex4-1}
\usepackage{color}
\usepackage{epstopdf}
\usepackage{graphicx}
\usepackage{dcolumn}
\usepackage{bm}
\usepackage{amssymb}   
\usepackage{amsmath}
\usepackage{braket}
\usepackage{csquotes}
\usepackage{hyperref}

\begin{document}

\preprint{APS/123-QED}

\title{Delayed sudden death of entanglement at exceptional points}

\author{Subhadeep Chakraborty}
 \email{c.subhadeep@iitg.ac.in}
\author{Amarendra K. Sarma}%
 \email{aksarma@iitg.ac.in}
\affiliation{
 Department of Physics, Indian Institute of Technology  Guwahati, Guwahati-781039, Assam, India}


%

\date{\today}

\begin{abstract}
Almost a decade ago, physicists encountered a strange quantum phenomenon that predicts an unusual death of entanglement under the influence of local noisy environment,known as entanglement sudden death (ESD). This could be an immediate stumbling block  in realizing all the entanglement based quantum information and computation protocols. In this work,  we propose a scheme to tackle such shortcomings by exploiting the phenomenon of exceptional points (EP). Starting with a binary mechanical $\mathcal{PT}$ symmetric system, realized over an optomechanical platform, we show that a substantial delay in ESD can be achieved via pushing the system towards an exceptional point. This finding has been further extended to higher (third) order exceptional point by  considering a more complicated tripartite entanglement into account.  

\begin{description}

\item[PACS numbers]
\end{description}
\end{abstract}

\maketitle
\section{\label{intro}Introduction}

$\mathcal{PT}$ symmetric quantum mechanics, as an extension of standard quantum theory into the complex domain, was introduced by Carl Bender and Stefan Boettcher in 1998. Followed by their seminal papers \cite{Bender1,Bender2}, this whole new class of non-Hermitian Hamiltonians was established that could exhibit a spectrum of entirely real eigenvalues under the only restriction $[H,\mathcal{PT}]=0$ \cite{rev}. Here, $\mathcal{P}$ refers to the parity operator that simply interchanges two of the constituent modes of the system, while $\mathcal{T}$ is the time-reversal operator that takes $i\rightarrow -i$. A more striking feature of such Hamiltonians is the breaking of $\mathcal{PT}$ symmetry, in which the eigen-spectra switches from being entirely real to completely imaginary. Such abrupt $\mathcal{PT}$ phase transition is marked by the presence of an exceptional point  \cite{heiss,ruter} where two (or more) eigenvalues and their corresponding eigenvectors coalesce and become degenerate.

While the search for $\mathcal{PT}$ symmetric devices is on, it occurs that one can easily implement such notions by judiciously providing gain and loss to an optical system. This leads to a remarkable exploration of $\mathcal{PT}$ phase transitions in particular to photonic systems, such as optical waveguides \cite{markis,musslimani,guo,ruter}, lattices and resonators \cite{regensburger,peng,feng,hodaei}. Moreover, based on such realizations, the existence of EP has further triggered many exotic phenomena which include nonreciprocal light propagation \cite{peng}, laser mode control \cite{feng,hodaei,hodaei2,peng2}, unidirectional invisibility \cite{regensburger,lin,longhi,feng2}, optical sensing \cite{wiersig,hodaei3,chen}, light stopping \cite{goldzak} and structuring \cite{miao}.  

At this point, one must note that most of these studies explored so far are confined in the so-called classical regime, as the involved components are essentially macroscopic in nature. Therefore, any $\mathcal{PT}$ symmetric device whose dynamics is governed by an intrinsic quantum mechanical equation of motion would provide a better insight into this theory. Along this direction, researchers have a proposed few architectures that include cold atoms \cite{hang,haag}, Bose-Einstein condensates \cite{cartarius}, optomechanical devices \citep{jing,xu,lu}, and recent circuit QED systems \cite{quijandria}. These quantum $\mathcal{PT}$ symmetric devices facilitate us to explore many intrinsic quantum properties, such as critical phenomena \cite{ashida}, entanglement \cite{chen2}, chiral population transfer \cite{xu2,doppler}, decoherence dynamics \cite{gardas}, and, information retrieval and criticality \cite{kawabata}. However, the true quantumness of a $\mathcal{PT}$ symmetric device still remains questionable, as while dealing with such gain (amplifying) and cooling (dampening) mechanisms one often abandons the associated quantum noises which rather must exist to preserve the proper commutation relation. So far, $\mathcal{PT}$ symmetry including quantum noises has been attempted in a very few studies \cite{agarwal,kepesdis,clerk,zhang}, which indicate a drastic difference from the usual (without considering quantum noises) predictions. Notably, in Ref \cite{ghamsari}, it has been shown the continuous variable (CV) entanglement, generated in a system of two coupled waveguides is seriously affected owing to the presence of quantum noises. Recently, incorporating gain saturation, it is proposed to reduce the influence of quantum noise \cite{ghamsari2} on entanglement.  

Entanglement \cite{horodecki}, being a form of correlation that is inherent to quantum systems, has become an invaluable resource for futuristic quantum computation and communication protocols. However, for a real-world implementation of such schemes, the longevity of the available entanglement is what experimentalists mostly concerned about. As, it is now well understood that any unavoidable interaction with external environment brings noise to the system which is substantially detrimental to the generated entanglement. One of such destructive manifestations, is the entanglement sudden death (ESD) \cite{esd} where the system losses entanglement in finite time. This unfortunate fate of entanglement has been both theoretically predicted and experimentally verified in a wide variety of entangled pairs involving atoms \cite{atomic}, photons \cite{photonic}, spin chains \cite{spin} and continuous Gaussian states \cite{gauss,gauss_expt}. Therefore, any manipulation that either avoids or delays ESD will help in executing various quantum information processing (QIP) protocols, that would otherwise be spoiled by the short entanglement lifetime. To overcome such shortcomings, a number of distinctive proposals have been put forward, such as quantum error correction \cite{error1,error2}, dynamical decoupling \cite{decoup1,decoup2}, decoherence free subspace \cite{decoher1,decoher2}, the quantum Zeno effect \cite{zeno1,zeno2}, delayed choice of decoherence suppression \cite{delay}, and quantum measurement reversal \cite{reverse1,reverse2,reverse3}. 

In this paper, we investigate such phenomenon of death of entanglement under the $\mathcal{PT}$ symmetric scenario. To the best of our knowledge, entanglement in $\mathcal{PT}$ symmetric geometries has been mostly dealt with optical binary systems, around the canonical $\mathcal{PT}$ phase transition point (EP). Therefore, it is intriguing ask: what happens to the entanglement evolution if one goes beyond the standard bipartite model to a more involved multipartite configuration, possessing higher-order exceptional point? To answer this question, we respectively consider a binary and a ternary mechanical $\mathcal{PT}$ symmetric system, where the mechanical gain and loss are induced in an optomechanical manner. Interestingly, we find that in both the cases the entanglement death can be substantially slowed down in the close vicinity of the exceptional points. This conclusion is further supported by analyzing the time evolution of two-mode Wigner mode functions near the EP. Finally, we discuss the effect of thermal noise on the bipartite and the tripartite entanglement evolutions.
\begin {figure}[t]\label{sys1}
\begin {center}
\includegraphics [width =5.5cm]{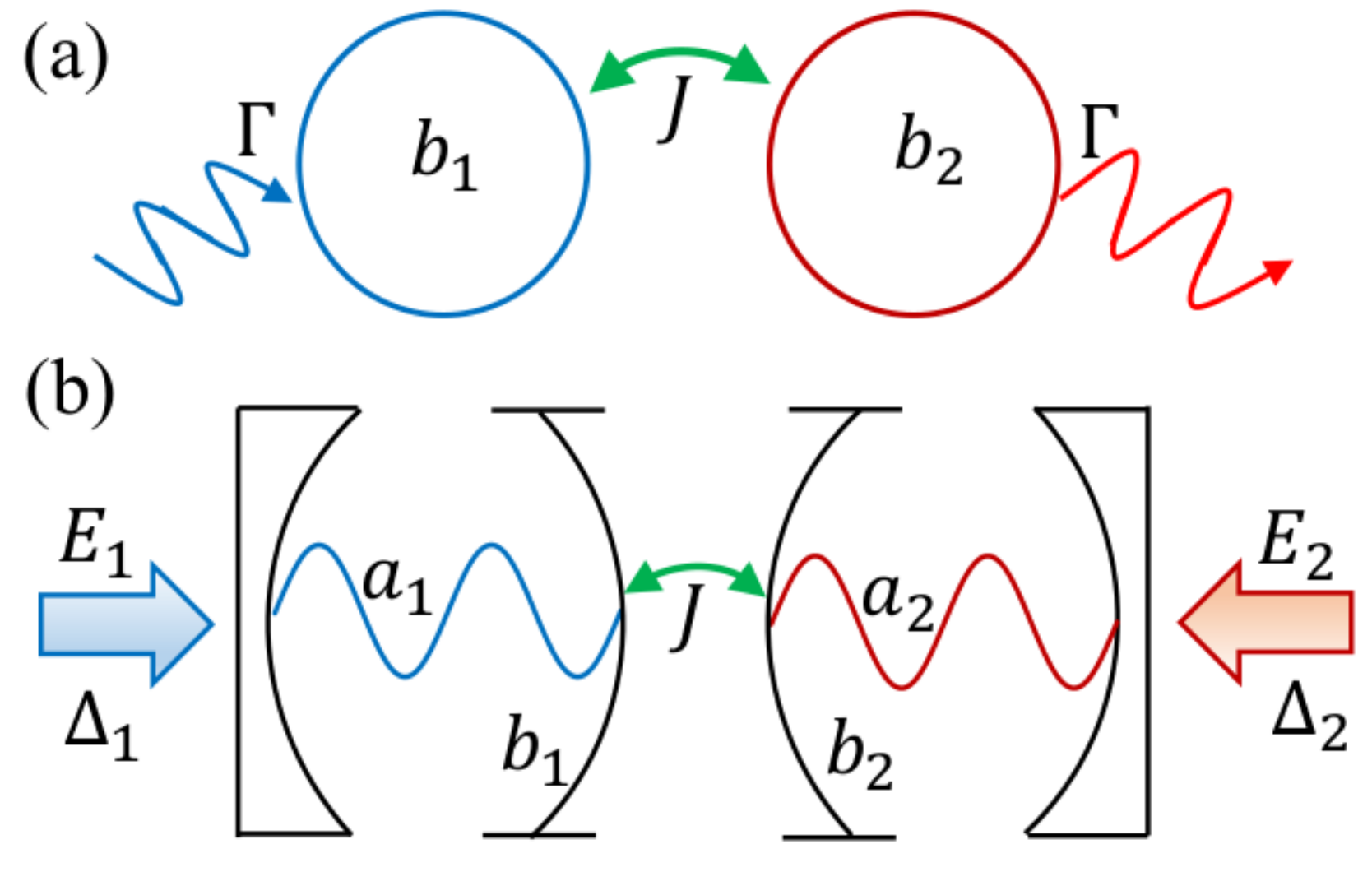}
\caption{\label {sys1}(Color online) (a) Schematic diagram of binary mechanical $\mathcal{PT}$ symmetric resonators, with optomechanically induced gain-loss. (b) Scheme for engineering mechanical gain/loss in an optomechanical platform. Here, two optomechanical cavities, respectively, driven at Stokes and Anti-Stokes of the driving lasers, are coupled mutually via a mechanical interaction.}
\end{center}
\end{figure}

\section{Entanglement in $\mathcal{PT}$ symmetric binary system}

Let us first consider a binary mechanical system as depicted in Fig. 1(a). This system essentially consists of two identical mechanical resonators, that are characterized by a mechanical frequency (damping rate) $\omega_m$ ($\gamma
$), and, are mutually coupled via a mechanical coupling of strength $J$. In addition, to induce the mechanical gain and loss into the system, we couple each of these resonators to an optomechanical cavity \cite{opto}, and, tune the respective cavity detunings to the Stokes and Anti-Stokes sidebands of the driving lasers. A schematic of this combined optomechanical system is illustrated in Fig. 1(b). Following the prescription as detailed in appendix, one can then write the effective dynamical equations corresponding to each mechanical resonators, as follows
\begin{subequations}\label{PT_eq}
\begin{align}
\dot{b_1}&=\left(\frac{\Gamma}{2}-\frac{\gamma}{2}\right) b_1+iJb_2+i\sqrt{\Gamma}a^{in^\dagger}_1+\sqrt{\gamma}b^{in}_1,\\
\dot{b_2}&=-\left(\frac{\Gamma}{2}+\frac{\gamma}{2}\right) b_2+iJb_1+i\sqrt{\Gamma}a^{in}_2+\sqrt{\gamma}b^{in}_2.
\end{align}
\end{subequations}
Here, $b_1,b_2$ ($b^\dagger_1,b^\dagger_2$) respectively be the annihilation (creation) operators of the gain and lossy resonators. $\Gamma$ is the optomechanically induced gain/loss rate, and, $a^{in}_j$ and $b^{in}_j$ ($j=1,2$) are the corresponding vacuum input noises, respectively, acting on the cavity fields and the mechanical resonators. In the parameter regime, where the effective optomechanical coupling is strong enough, one can further discard the intrinsic mechanical damping as $\Gamma\gg\gamma$. However, to simulate the actual physical condition, we retain a finite damping $\gamma>0$ throughout our calculations. 

By ignoring the quantum noises, one can recast the above equation as $\dot{u}(t) = -i H_2u(t)$. Here, $u^T(t)=\left(q_1(t),p_1(t),q_2(t),p_2(t)\right)$ is the state vector, written in terms of the dimensionless CV quadrature operators $q_j\equiv\left(b_j+b^\dagger_j\right)/\sqrt{2}$ and  $p_j\equiv\left(b_j-b^\dagger_j\right)/i\sqrt{2}$ (with $j=1,2$), while $H_2$ be the non-Hermitian Hamiltonian, given by
\begin{align}
 H_2=i\left(
 \setlength\arraycolsep{4 pt}
 \begin{array}{cccc}
 \frac{\Gamma}{2} & 0 & 0 &-J\\[2 pt]
 0 & \frac{\Gamma}{2} & J & 0\\[2 pt]
 0 & -J & -\frac{\Gamma}{2} & 0\\[2 pt]
 J & 0 & 0 & -\frac{\Gamma}{2}
 \end{array}
 \right).
 \end{align}
One can then easily verify that under the simultaneous $\mathcal{PT}$ operation, the Hamiltonian remains invariant, i.e., $\left[\mathcal{PT},H_2\right]=0$

Now, in order to study the $\mathcal{PT}$ phase transition, we first diagonalize the Hamiltonian and find the eigenfrequencies:
\begin{equation}
\omega_{\pm}=\pm\sqrt{J^2-\left(\frac{\Gamma}{2}\right)^2}.
\end{equation}
It is evident that respectively for $J>\Gamma/2$ and $J<\Gamma/2$ there exist two distinct phases: one that includes all the real eigenvalues, namely a $\mathcal{PT}$ symmetric phase, while the other possesses purely imaginary eigenvalues, namely a broken $\mathcal{PT}$ symmetric phase. The critical coupling strength which separates these two phases is located at $J_{c_2}=\Gamma/2$. An exact situation can be found in Fig. 2, where we depict the eigenfrequencies as a function of the normalized coupling strength $J/\Gamma$. Also, we observe that this same $J_{c_2}$ marks the exceptional point (of order 2) of this system where the two eigenvalues coalesce. 

\begin {figure}[t]\label{pt2}
\begin {center}
\includegraphics [width =9cm]{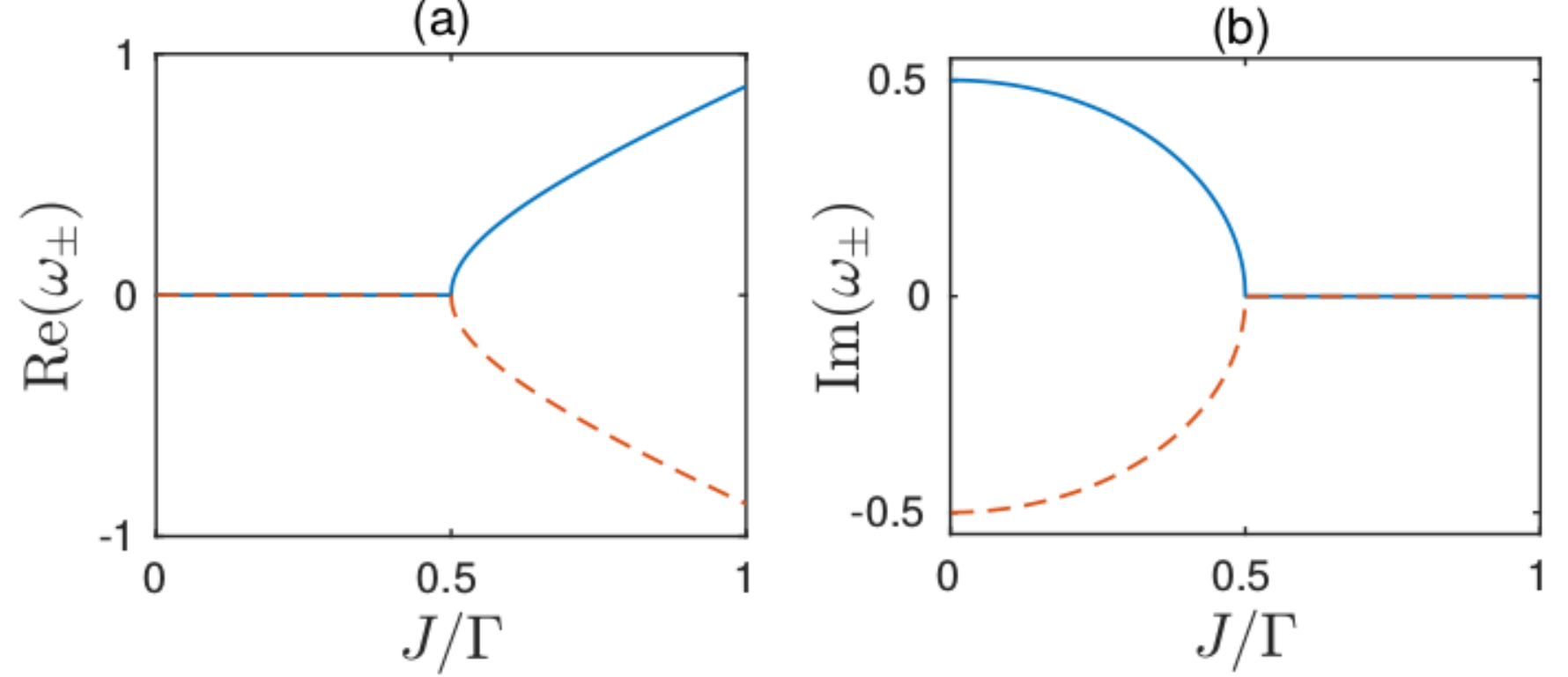}
\caption{\label {pt2}(Color online) (a) The real, and, (b) the imaginary parts of $\omega_\pm$ \textit{vs}. the normalized coupling coefficient $J/\Gamma$. The system exhibits an exceptional point (of order 2) at $J=\frac{\Gamma}{2}$. }
\end{center}
\end{figure}
Next, we take the full quantum noises into the account and rewrite Eq. \eqref{PT_eq} in a more compact from: $\dot{u}(t) = Au(t) + n(t)$. Here, $u(t)$ be the same CV state vector, $A=-iH_2$ is the drift matrix, and, $n^T(t)=\left(\sqrt{\Gamma} Y^{in}_1 + \sqrt{\gamma}Q^{in}_1, \sqrt{\Gamma}X^{in}_1+\sqrt{\gamma}P^{in}_1, -\sqrt{\Gamma}Y^{in}_2 \right. $\hspace{0.01cm}$\left. +\sqrt{\gamma}Q^{in}_2, \sqrt{\Gamma}X^{in}_2+\sqrt{\gamma}P^{in}_2\right)$ is the matrix of corresponding noises. The input noise quadratures, as used in $n^T(t)$, are respectively defined as follows: $X^{in}_c \equiv \left(a^{in}_c + a^{in^\dagger}_c\right)/\sqrt{2}, Y^{in}_c \equiv\left(a^{in}_c - a^{in^\dagger}_c\right)/i\sqrt{2}$, and, $Q^{in}_c \equiv \left(b^{in}_c + b^{in^\dagger}_c\right)/\sqrt{2}, P^{in}_c \equiv\left(b^{in}_c - b^{in^\dagger}_c\right)/i\sqrt{2}$ where $c=1,2$. A stable solution of this equation requires all the eigenvalues with negative real parts. In what follows, we find that one can have such solutions only when the system remains in the $\mathcal{PT}$ symmetric phase.

We further note that owing to the above linearized dynamics and zero-mean Gaussian nature of the quantum noises, the system retains its Gaussian characteristics. In turn, one can adopt the standard covariance matrix (CM) formalism to fully describe the system \cite{covar}. Let $V(t)$ be the CM with each elements defined as $V_{ij}(t)=\langle u_i(t)u_j(t) + u_j(t)u_i(t) \rangle/2$. Then, one has the following equation of motion as satisfied by the CM 
\begin{equation}\label{vdot}
\dot{V}(t)=AV(t)+V(t)A^T+D.
\end{equation}
Here, $D=\left(\frac{\Gamma}{2}+\gamma\left(n_{th}+\frac{1}{2}\right)\right)\mathrm{diag}\left(1,1,1,1\right)$ is the matrix of the noise correlations, obtained under the Markovian assumption and $\langle n_i(t)n_j(t^\prime)+n_j(t^\prime)n_i(t)\rangle/2=\delta\left(t-t^\prime\right)D_{ij}$. 

The above Eq. \ref{vdot} is an inhomogeneous first  order differential equation which can be solved numerically with a proper initial condition. As studying the quantum correlation is of our prime concern, here, we consider this input state to be a generic CV entangled state, i.e., a two mode squeezed state $|z\rangle = e^{r\left(b^\dagger_1b^\dagger_2-b_1b_2\right)}|0,0\rangle$ with $r$ being the squeezing parameter. Finding a solution of Eq. \eqref{vdot}, leads us to write $V(.)$ as
\begin{align}
V\equiv\left(
  \begin{array}{cc}
    A & C \\
    C^T & B \\
  \end{array}
\right),
\end{align}
where $A$, $B$ and $C$ are $2\times2$ block matrices, respectively, corresponding to the local covariance matrices of resonators $1$ and $2$, and the nonlocal correlation between them. One can, then, gauge the degree of quantum entanglement by calculating the so-called logarithmic negativity $E_N$,  defined as:
$E_N=\mathrm{max}\left[0,-\ln2\nu^-\right]$ \cite{entg1,entg2}.
Here $\nu^-\equiv2^{-1/2}\left[\Sigma(V)-\sqrt{\Sigma(V)^2-4\mathrm{det}V}\right]^{1/2}$ is the smallest symplectic eigenvalue of the partial transpose of $V$ with $\Sigma(V)\equiv \mathrm{det}(A)+\mathrm{det}(B)-2\mathrm{det}(C)$.  
\begin {figure}[t]\label{pt2_en}
\begin {center}
\includegraphics [width =9cm,height=4cm]{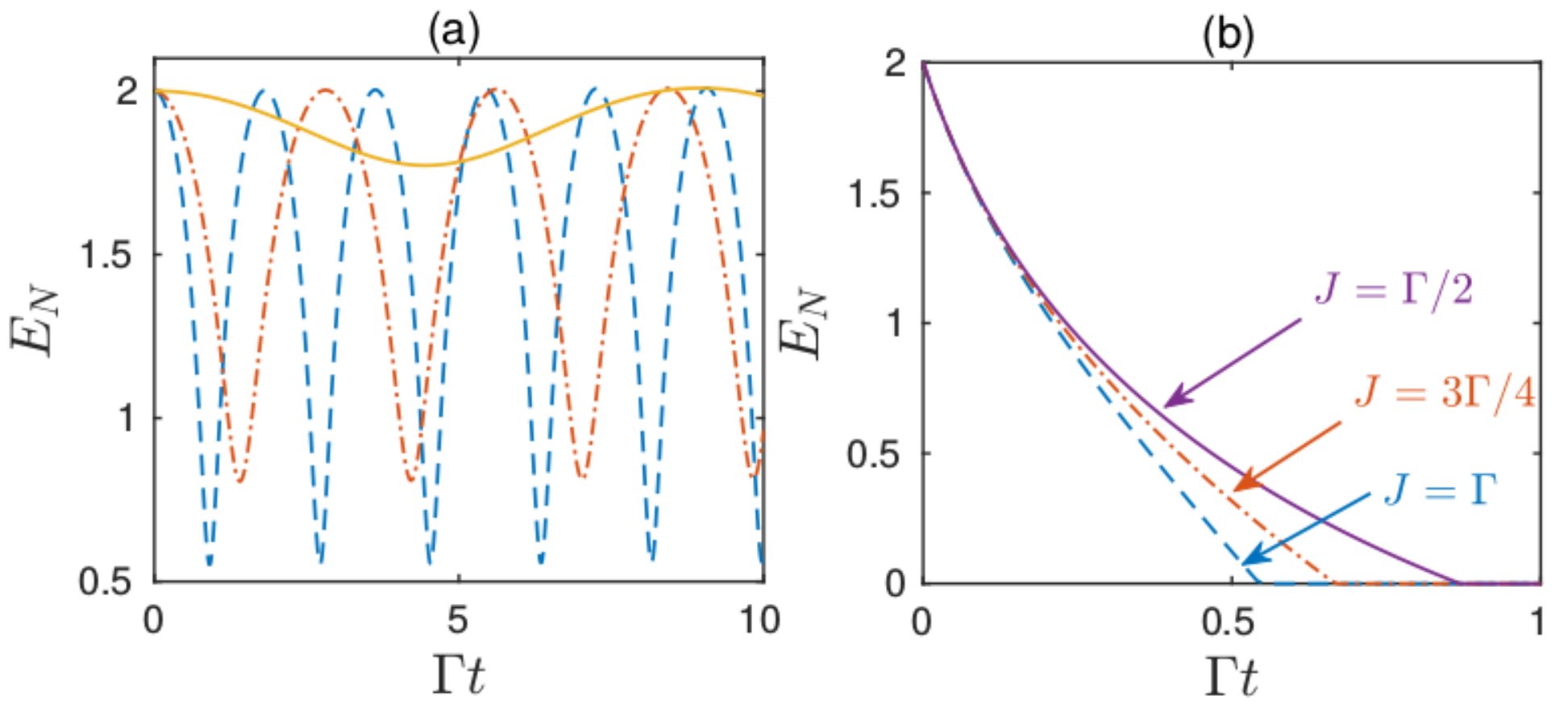}
\caption{\label {pt2_en}(Color online) Entanglement evolution between the gain-loss resonators, (a) in the absence and (b) in the presence of noises. The parameters used are $\Gamma=1$ and $\gamma=10^{-3}$. In panel (a) the blue (dashed), red (dash-dotted) and yellow (solid) lines respectively corresponds to the $J=\Gamma$, $J=0.75\Gamma$ and $J=0.53\Gamma$.}
\end{center}
\end{figure}

In Fig. 3(a) we first show the time evolution of the quantum entanglement in the absence of any noises. One can see that when the system is in the $\mathcal{PT}$ symmetric phase, the entanglement oscillates periodically. This oscillations could be attributed to the nature of the eigenvalues $\pm i \sqrt{J^2-J^2_{c_2}}$ of $A$, as obtained for $J>J_{c_2}$. However, as we approach the EP, we notice a lesser oscillation with a longer time period. Finally, in the close vicinity of EP $\left(J\rightarrow J_{c_2}\right)$, the entanglement dynamics almost \enquote{freezes out}, i.e., it takes a longer time to complete one oscillation. In Fig. 3(b), we show the same entanglement evolution, but, now taking the quantum noises taking into the account. One can see that as soon as noise is introduced into the system, the entanglement quickly decays to zero, a typical ESD like behavior. However, a more notable feature of this figure is the delayed death of entanglement, as achieved via pushing the system towards the exceptional point (EP).

To further support this conclusion, in Fig. 4 we compare the two-mode Wigner functions for two different coupling strengths, at two different times. One can see that at $\Gamma t=0.5$ both these ${W(q_1,q_2)^{p_1 =0}_{p_2=0}}$ exhibit a squeezing like behavior which is sufficient to ensure the presence of entanglement between these two-modes. However, at a subsequent time $\Gamma t=0.75$, one finds the ${W(q_1,q_2)^{p_1 =0}_{p_2=0}}$ corresponding $J=\Gamma$ almost losses its squeezing characteristic, while the other still retains it with a finite degree. Moreover, it is also important to note that for both these coupling strengths, the Wigner functions remain localized in the phase-space. It means that there is no abrupt stretching associated with ${W(q_1,q_2)^{p_1 =0}_{p_2=0}}$ that could push the system towards instability.  
\begin {figure}[t]\label{wig}
\begin {center}
\includegraphics [width =9cm]{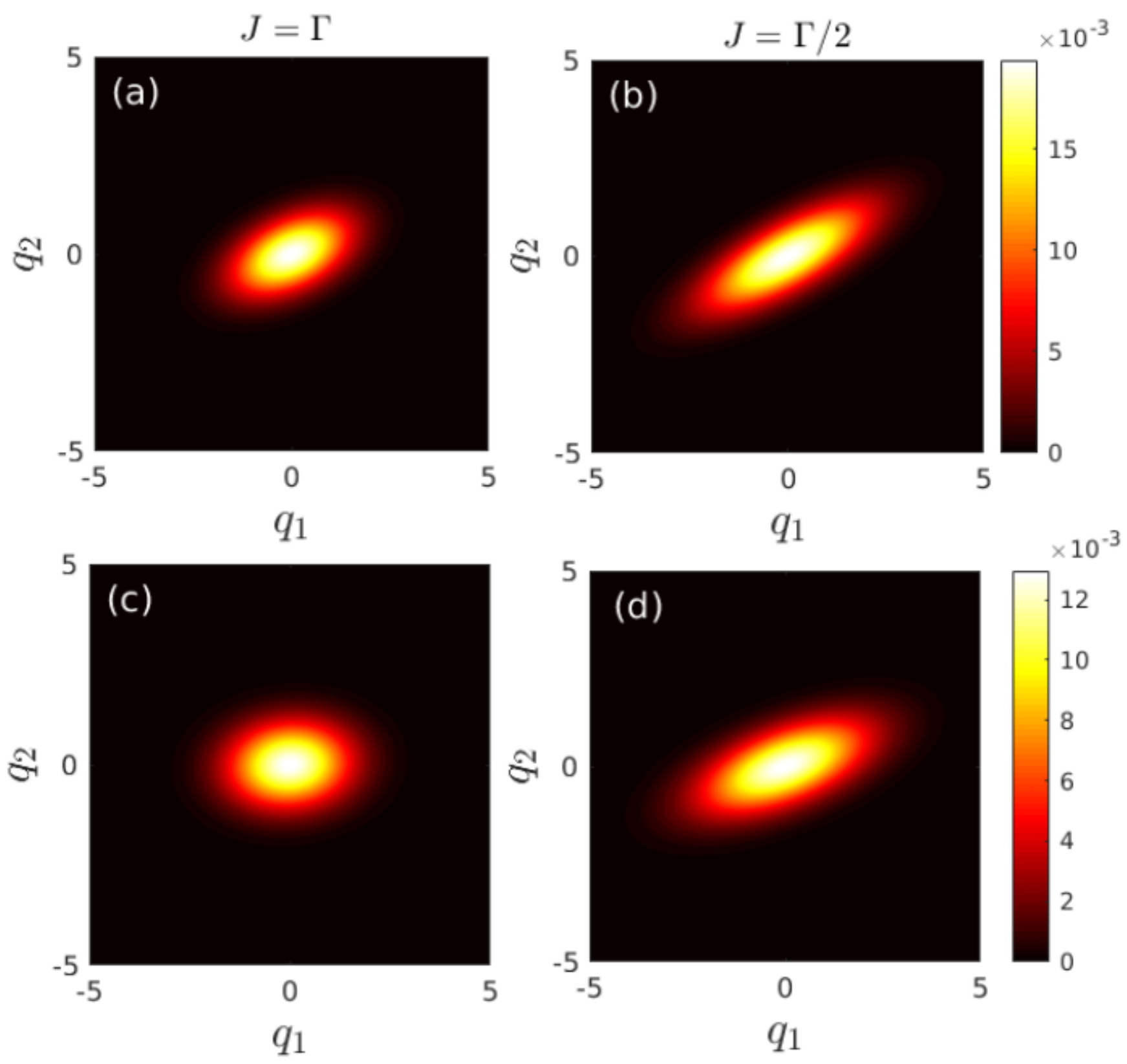}
\caption{\label {wig}(Color online) Time evolution of Wigner functions ${W(q_1,q_2)^{p_1 =0}_{p_2=0}}$, respectively, at $\Gamma t=0.5$ (upper two panels), and $\Gamma t=0.75$ (bottom two panels).}
\end{center}
\end{figure}

\section{Entanglement in $\mathcal{PT}$ symmetric ternary system}

Motivated by these results, we now extend this same strategy to the higher-order exceptional points. A possible realization that supports a EP3 (exceptional point of order 3) would be a ternary mechanical system where the gain and lossy resonators are being separated by a neutral one (see Fig. 5). Then, proceeding in a similar manner, one can write the following equation of motion, as satisfied by the each mechanical resonators 
\begin{subequations}\label{PT_eq3}
\begin{align}
\dot{b_1}&=\left(\frac{\Gamma}{2}-\frac{\gamma}{2}\right) b_1+iJb_2+i\sqrt{\Gamma}a^{in^\dagger}_1+\sqrt{\gamma}b^{in}_1,\\
\dot{b_2}&=-\frac{\gamma}{2} b_2+iJb_1+iJb_3+\sqrt{\gamma}b^{in}_2,\\
\dot{b_3}&=-\left(\frac{\Gamma}{2}+\frac{\gamma}{2}\right) b_3+iJb_2+i\sqrt{\Gamma}a^{in}_2+\sqrt{\gamma}b^{in}_3.
\end{align}
\end{subequations}
Here, $b_1,b_2,b_3$ ($b^\dagger_1,b^\dagger_2,b^\dagger_3$),  refer to the annihilation (creation) operators of the gain, neutral and lossy resonators, respectively, while the other parameters remain unchanged with the previous descriptions. Then in the absence of any noises, one can have the following cubic algebraic equation as obeyed by each of these eigenfrequencies $\omega_n$ $\left(n\in\{-1,0,1\}\right)$
\begin{equation}
\omega_n \left(\omega^2_n + \frac{\Gamma^2}{4} - 2 J^2 \right) =0.
\end{equation}
It is evident that for a critical coupling strength $J_{c_3}=\frac{\Gamma}{2\sqrt{2}}$, all the three eigenfrequencies merge at $\omega_n=0$. This characteristic feature of EP3 has been pictorially demonstrated in Fig. 6, where we show the dependence of the eigenfrequencies on the normalized coupling strength $J/\Gamma$. 
\begin {figure}[!t]\label{sys2}
\begin {center}
\includegraphics [width =6.5cm]{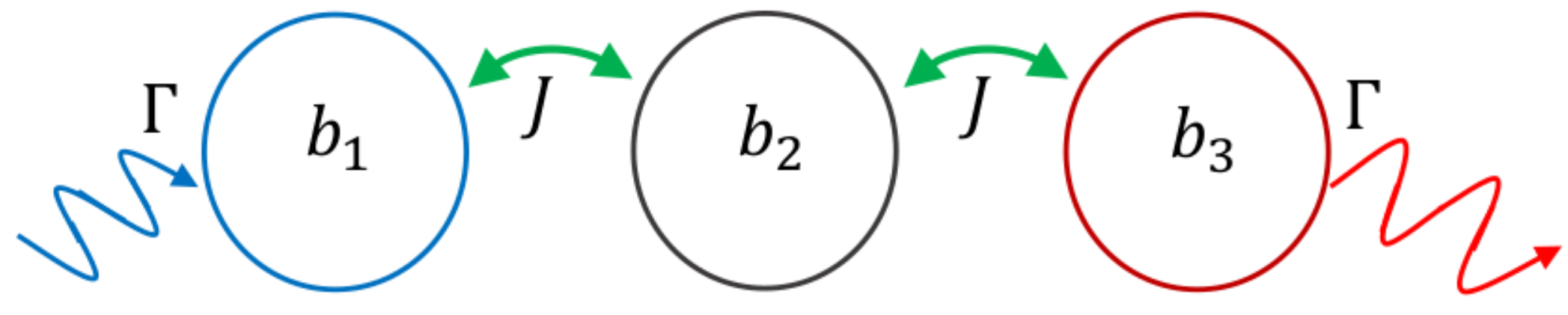}
\caption{\label {sys2}(Color online) Schematic illustration of $\mathcal{PT}$ symmetric ternary mechanical setup. Here, the side resonators are subjected to equivalent gain and loss, while the middle one is neutral.}
\end{center}
\end{figure}
\begin {figure}[t]\label{pt3}
\begin {center}
\includegraphics [width =9cm]{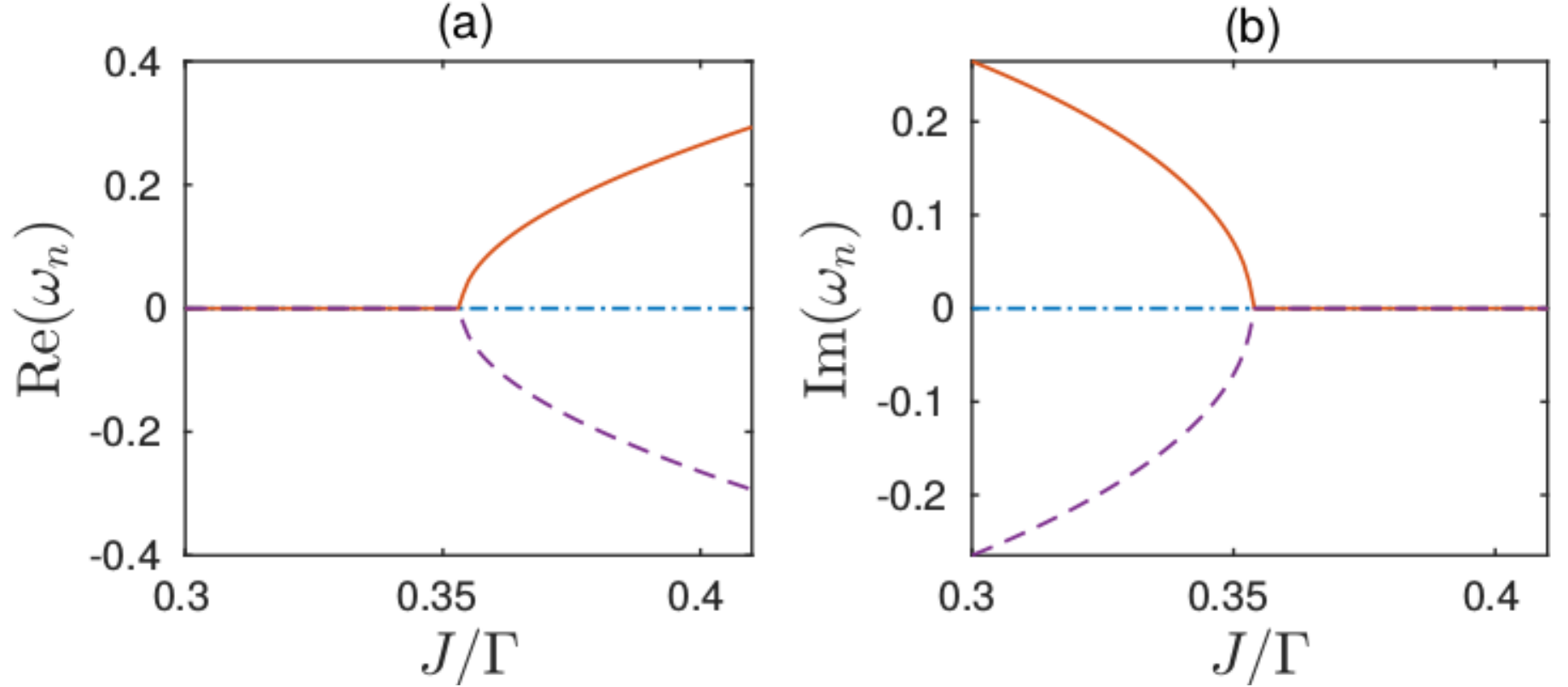}
\caption{\label {pt3}(Color online) (a) The real and (b) the imaginary parts of $\omega_n$ as a function of the normalized coupling coefficient $J/\Gamma$. The three eigenfrequencies merge at $J=\frac{\Gamma}{2\sqrt{2}}$, exhibiting a third-order exceptional point}
\end{center}
\end{figure}

Now, to take this discussion further ahead, we consider the case of tripartite entanglement between theses three mechanical resonators, under the $\mathcal{PT}$ symmetric scenario. The measurement of such complicated multipartite entanglement goes as follows. Let us first define the following linear combinations of the quadrature variances:
\begin{subequations}  
\begin{align}
x\equiv& h_1 q_1 + h_2 q_2  + h_3 q_3,\\
y\equiv& g_1 p_1 + g_2 p_2  + g_3 p_3,
\end{align}
\end{subequations}
where $q_1,q_2,q_3$ $\left(p_1,p_2,p_3\right)$ be the position (momentum) operators of the gain, neutral and lossy resonators, respectively, while $h_k$ and $g_k$ are being any arbitrary real parameters. Then, following the prescription as proposed in Refs. \cite{lock, reid, gonza}, we employ the nonseparability measurement
\begin{equation}
S = \langle \left(\Delta x\right)^2 \rangle + \langle \left(\Delta y\right)^2 \rangle.
\end{equation}
The three-party state is then said to be \textit{genuinely tripartite entangled} iff it negates the following single inequality
\begin{eqnarray}\label{tri}
S &&\geq \mathrm{min}\{|h_3g_3|+|h_1g_1+h_2g_2|,\\ \nonumber
&&|h_2g_2|+|h_1g_1+h_3g_3|,|h_1g_1|+|h_2g_2+h_3g_3|\},
\end{eqnarray}
whereas, only the violation of any of theses inequalities
\begin{equation} \label{sepa}
S\geq\left(|h_kg_k|+|h_lg_l+h_mg_m|\right),
\end{equation}
for a given permutation of $\left\{k,l,m\right\}$ of $\left\{1,2,3\right\}$, guarantees a \textit{full tripartite inseparability}. These classes of nonseparabilities may sound equivalent, but as pointed out by Teh and Reid \cite{reid} they remain indistinguishable only for pure quantum states, while for mixed states meeting the full inseparability criteria does not suffice to claim multipartite entanglement. 

In order to numerically evaluate the bounds of Eq. \ref{tri} and \ref{sepa}, we now consider the following set of the parameters $h_1=g_1=1$ and $g_2=g_3=-h_2=-h_3=1/\sqrt{2}$. The reason behind such particular combination is $\left[q_1-(q_2+q_3)/\sqrt{2},p_1+(p_2+p_3)/\sqrt{2}\right]=0$, which allows to have an arbitrary good degree of violation of \ref{tri}. Then, one has to fulfill $S<1$ to ensure the emergence of genuine tripartite entanglement, while $S<2$ will suffice to confirm at least full tripartite inseparability. 
\begin {figure}[t]\label{en_pt3}
\begin {center}
\includegraphics [width =6.5cm,height = 5cm]{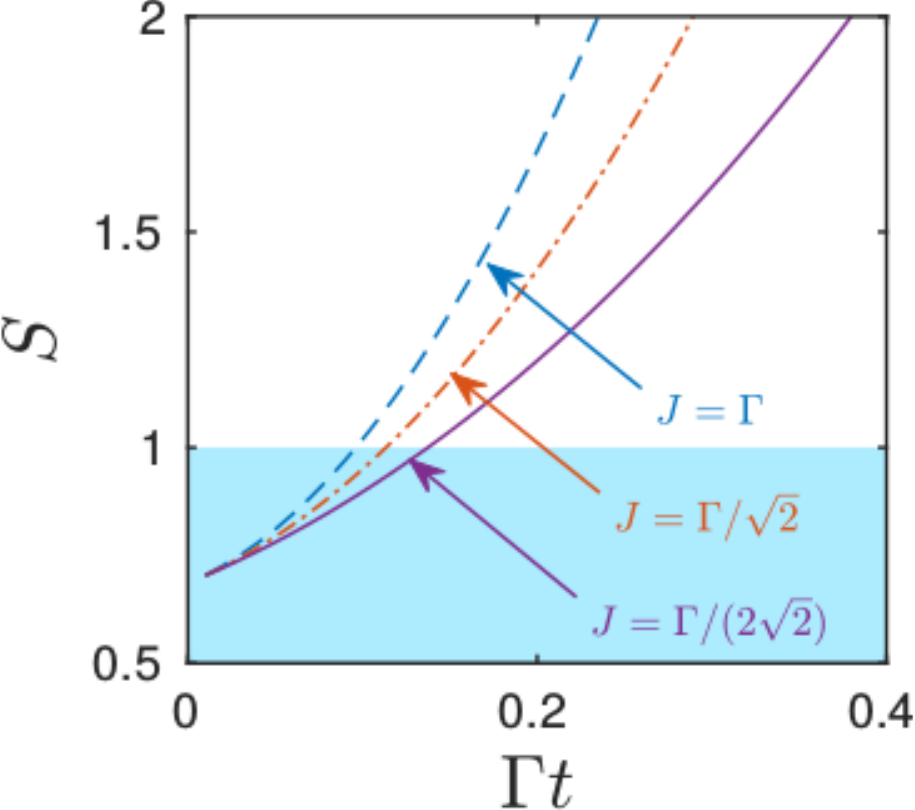}
\caption{\label {en_pt3}(Color online) Time evolution of the nonseparability criteria $S$, for three different coupling strengths. The shaded area guarantees the presence of genuine tripartite entanglement. The parameters used are same with Fig. 3.}
\end{center}
\end{figure}

Fig. 7 depicts the time evolution of $S$, starting from a CV GHZ state with the squeezing parameters $r_1, r_2$ being equal to 1 \cite{lock}. It is observed that, with the inclusion of quantum noise, the tripartite state quickly suffers a sudden death of (genuine tripartite) entanglement, followed by a fully tripartite inseparable state. However, it is remarkable to note that at EP3, such three-party state losses entanglement more slowly as compared to any other points in unbroken the $\mathcal{PT}$ symmetric phase. One may also notice that by operating near the EP3, it is possible to prolongate the time pertaining to a full tripartite inseparable state.

Finally, in Fig. 8 we examine the effect of thermal noise on the bipartite (Fig. 8(a)) and tripartite entanglement (Fig. 8(b)) evolution. As expected, the degradation of quantum entanglement with increasing thermal phonon is observed. However, it is worthwhile to note that, the available bipartite (tripartite) entanglement achieved in the close vicinity of the EPs (EP3) is fairly robust. This makes our proposed scheme a viable means to improve the entanglement sustainability with the aid of exceptional points. 
\begin {figure}[t]\label{en_pt3}
\begin {center}
\includegraphics [width =9cm]{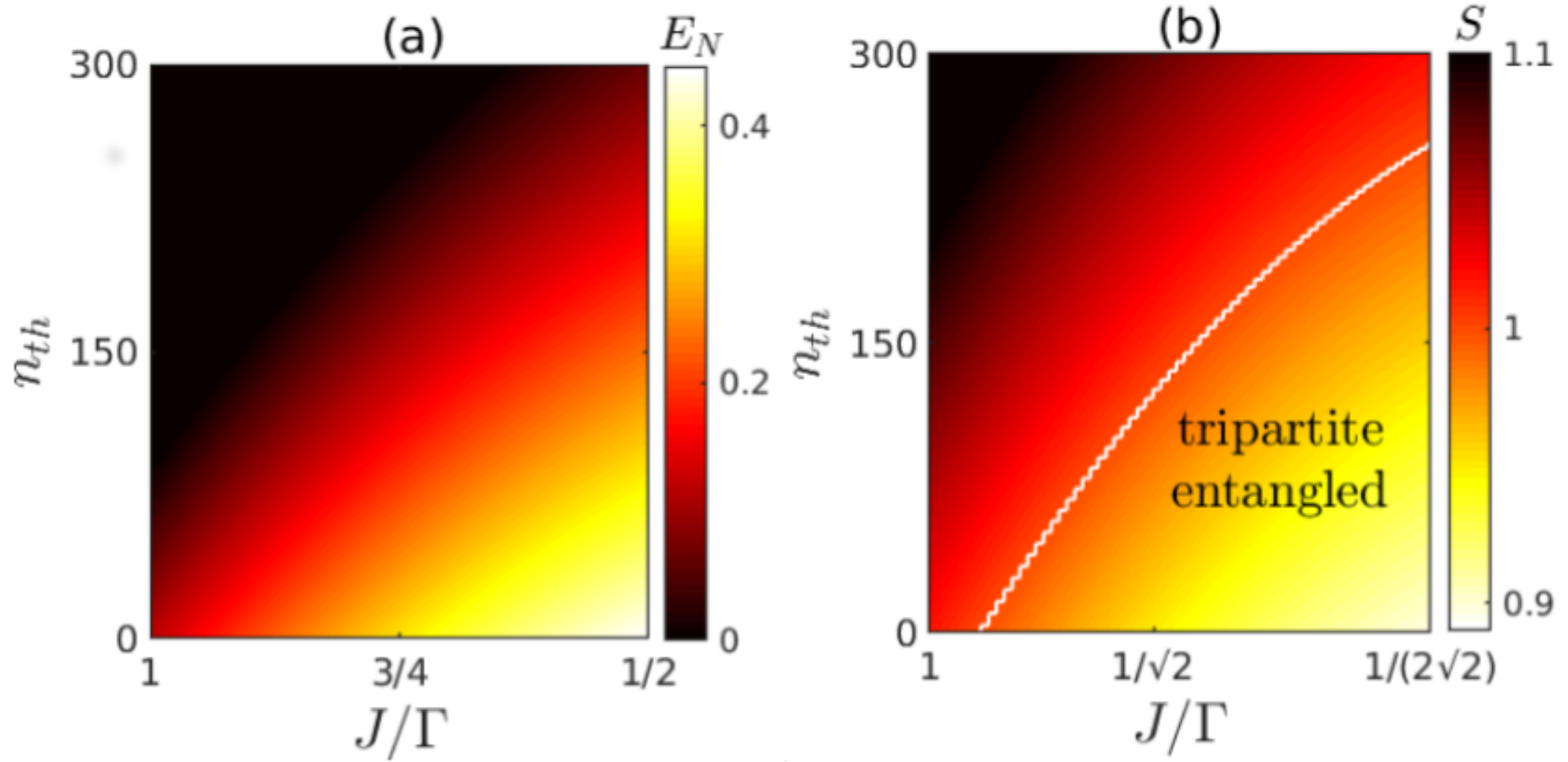}
\caption{\label {en_pt3}(Color online) (a) Bipartite and (b) tripartite entanglement measures, as a function of the normalized coupling $J/\Gamma$ and the number of thermal phonons $n_{th}$. The time corresponding to each snapshots are, respectively, (a) $\Gamma t=0.5$ and (b) $\Gamma t=0.1$. In Fig. 8(b), the white line separates between two distinct regimes of genuine tripartite entanglement and full tripartite inseparability.}
\end{center}
\end{figure}

\section{Conclusion}

In conclusion, we have performed a systematic study to unravel the relation between the death of entanglement and the phenomenon of exceptional points. The architectures that we have specifically focused on, are essentially, binary and ternary mechanical $\mathcal{PT}$ symmetric systems with optomechanically induced gain-loss. Our study shows that a substantial delay in the sudden death of entanglement is possible at exceptional points. It is also shown that near the EPs, the available entanglement survives to a much higher degree of thermal phonons. These findings may pave the ways for exploiting $\mathcal{PT}$ symmetric devices as novel means to control the entanglement dynamics in different QIP protocols. It is also worthwhile to note that our approach could in principle, be applied in various photonic and phononic $\mathcal{PT}$ symmetric systems, with engineered gain-loss mechanisms. Further efforts along this directions may include the investigation of quantum entanglement in the broken $\mathcal{PT}$ symmetric regime, where the effect of gain saturation must be treated with care.

\section*{ACKNOWLEDGEMENT}

S. Chakraborty would like to acknowledge MHRD, Government of India for providing a financial support for his research.

\setcounter{section}{0}
\setcounter{subsection}{0}
\section*{\label{opto_app}appendix: Cavity optomechanics based architecture to realize the gain (loss) in a mechanical system}
\setcounter{equation}{0}
\renewcommand{\theequation}{A.\arabic{equation}}
In this section, we briefly outline the derivation of the optomechanically induced gain and loss in an mechanical system. To begin with, we first consider a generic cavity optomechanical system, consisting of a single cavity mode of frequency $\omega_c$ and a mechanical mode of frequency $\omega_m$. Following a rotating frame transformation at a (laser) frequency $\omega_l$, the Hamiltonian of this system reads ($\hbar=1$):
\begin{equation}
 H=\Delta_0 a^\dagger a+\omega_m b^\dagger b-g a^\dagger a(b^\dagger+b)+E_0 \left(a^\dagger +a\right),
\end{equation}
where $a$ ($a^\dagger$) and $b$ ($b^\dagger$) are, respectively, the annihilation (creation) operators of the cavity field and the mechanical resonator. $g$ is the strength of the single-photon optomechanical coupling, and, $E_0$ is the driving amplitude with $\Delta_0=\omega_c-\omega_l$ being the cavity detuning. Taking the fluctuation-dissipation processes into account, the dynamics of the system is then fully described by the following set of nonlinear quantum Langevin equations (QLEs),
\begin{subequations}
\begin{align}
\dot{a} &=-(i\Delta_0 +\kappa/2)a+i g a(b^\dagger + b)-i E_0 +\sqrt{\kappa}a^{in}, \\
\dot{b} &=-(i\omega_m+\gamma/2)b+i g a^\dagger a +\sqrt{\gamma}b^{in}.
\end{align}
\end{subequations}
Here $\kappa$ ($\gamma$) is the cavity decay (mechanical damping) rate. $a^{in}$ is the zero-mean vacuum input noise operator, satisfying the only non-zero correlation function: $\langle a_{in}(t)a^\dagger _{in}(t^\prime)\rangle=\delta(t-t^\prime)$,
and $b^{in}$ refers to the random Brownian noise operator, with zero-mean value and Markovian correlation functions, given by: $\langle b^{in^\dagger} (t) b^{in}(t^\prime) \rangle = n_{th}\delta(t-t^\prime)$, $\langle b^{in} (t) b^{in^\dagger} (t^\prime) \rangle = (n_{th}+1)\delta(t-t^\prime)$. The parameter $n_{th}=\left[\mathrm{ exp}\left(\frac{\hbar \omega_m}{k_BT}\right)-1\right]^{-1}$ denotes the mean thermal phonon number at temperature $T$ ($k_B$ being the Boltzmann constant).

For a strongly driven cavity, we now adopt the standard linearization technique and expand each operators as a sum of its $c$-number classical steady-state value plus a time-dependent zero-mean quantum fluctuation operator, i.e., $a(t)\rightarrow\alpha+a(t)$ and $b(t)\rightarrow\beta+ b(t)$. These steady states values could then be obtained by solving the following nonlinear algebraic equations:
\begin{subequations}
 \begin{align}
  (i\Delta+\kappa/2)\alpha+iE_0 &=0,\\
  (i\omega_m+\gamma_m/2)\beta-i g |\alpha|^2&=0,
 \end{align}
\end{subequations}
where $\Delta=\Delta_0-2g\mathrm{Re}(\beta)$ is the effective cavity detuning. On the other hand, the dynamics of the quantum fluctuations are given by the linearized QLEs (valid in the limit of $|\alpha|\gg1$), written as:
\begin{subequations}\label{fluc}
 \begin{align}
\dot{a} &=-(i\Delta+\kappa/2)a+i G(b^\dagger + b)+ \sqrt{\kappa}a^{in},\\
\dot{b} &=-(i\omega_m+\gamma/2) b+i G ( a^\dagger + a)+\sqrt{\gamma}b^{in},
 \end{align}
\end{subequations}
with $G=g|\alpha|$ being the effective many-photon optomechanical coupling strength.

Next, we introduce two slowly varying operators: $\tilde{a}=a e^{i\Delta t}$ and $\tilde{b}= b e^{i\omega_m t}$, and, rewrite Eq. \eqref{fluc} in the following way:
\begin{subequations}\label{tilde}
 \begin{align}
  \dot{\tilde{a}} &= -\frac{\kappa}{2}\tilde{a} +iG\left(\tilde{b}^\dagger e^{i(\Delta+\omega_m)t}+\tilde{b}e^{i(\Delta-\omega_m)t}\right)+\sqrt{\kappa}\tilde{a}^{in},\\
  \dot{\tilde{b}} &=-\frac{\gamma}{2}\tilde{b}+iG\left(\tilde{a}^\dagger e^{i(\omega_m+\Delta)t}+\tilde{a}e^{i(\omega_m-\Delta)t}\right)+\sqrt{\gamma}\tilde{b}^{in}.
 \end{align}
\end{subequations}
Note that in the above equations, we have also defined the new noise operators $\tilde{a}^{in}=a^{in}e^{i\Delta t}$ and $\tilde{b}^{in}=b^{in}e^{i\omega_m t}$ which possess the same correlation functions.

We now proceed to discuss how to realize gain (damping) in the mechanical resonator, in a pure quantum mechanical way. To do so, we first assume that the cavity is resonant with the Stokes sideband of the driving laser, $\Delta=-\omega_m$, and invoke the rotating wave approximation (RWA) (which is justified in the limit of $\omega_m\gg \{G,\kappa,\gamma\}$) to obtain:
\begin{subequations}
 \begin{align}
  \dot{\tilde{a}} &= -\frac{\kappa}{2}\tilde{a}+iG\tilde{b}^\dagger+\sqrt{\kappa}\tilde{a}^{in},\\
  \dot{\tilde{b}} &= -\frac{\gamma_m}{2}\tilde{b}+iG\tilde{a}^\dagger+\sqrt{\gamma}\tilde{b}^{in}.\label{b}
  \end{align}
\end{subequations}
Under the condition that the cavity decay rate is much larger than the effective optomechanical coupling strength, $\kappa\gg \{G,\gamma_m\}$,  one can adiabatically eliminate the cavity field and gets
\begin{equation}\label{ad}
 \tilde{a}=i\frac{2G}{\kappa} \tilde{b}^\dagger+\frac{2}{\sqrt{\kappa}}\tilde{a}^{in}.
\end{equation}
Following a substitution of Eq. \eqref{ad} in Eq. \eqref{b}, we end up with the following equation describing effective the dynamics of the mechanical resonator:
\begin{equation}\label{gain_eq}
\dot{\tilde{b}}=\left(\frac{\Gamma}{2}-\frac{\gamma}{2}\right) \tilde{b}+i\sqrt{\Gamma}\tilde{a}^{in^\dagger}+\sqrt{\gamma}\tilde{b}^{in}.
\end{equation}
Here, one should note the inclusion of the following two terms: $\Gamma=\frac{4G^2}{\kappa}$ which quantifies the amount of optomechanically induced gain in the mechanical resonator, and, the cavity induced noise term $\sqrt{\Gamma}a^{in^\dagger}$ which helps to preserve the right commutation relation. 

On the contrary, when the cavity is resonant with the Anti-Stoke sideband of the driving laser, $\Delta=\omega_m$, following a similar procedure, one obtains:
\begin{equation}\label{loss_eq}
\dot{\tilde{b}}=-\left(\frac{\Gamma}{2}+\frac{\gamma}{2}\right)\tilde{b}+i\sqrt{\Gamma}\tilde{a}^{in}+\sqrt{\gamma}\tilde{b}^{in},
\end{equation}
where the same $\Gamma=\frac{4G^2}{\kappa}$ corresponds to the optomechanically induced loss or damping in the mechanical resonator.

\end{document}